\documentclass{pasj01}
\Received{2016 March 4}%{$\langle$reception date$\rangle$}
\Accepted{2016 April 13}%{$\langle$acception date$\rangle$}
\Published{}%{$\langle$publication date$\rangle$}
\begin{document}

\title{The relationship between variable and polarized optical spectral components of luminous type~1 non-blazar quasars}
\author{Mitsuru \textsc{Kokubo} \altaffilmark{1} \altaffilmark{2}}%
\altaffiltext{1}{Department of Astronomy, School of Science, the University of Tokyo, 7-3-1 Hongo, Bunkyo-ku, Tokyo 113-0033, Japan}
 \altaffiltext{2}{Institute of Astronomy, the University of Tokyo, 2-21-1 Osawa, Mitaka, Tokyo 181-0015, Japan}
\email{mkokubo@ioa.s.u-tokyo.ac.jp}

\KeyWords{accretion, accretion disks --- galaxies: active --- galaxies: nuclei ---  polarization --- quasars: individual (B2~1208+32, Ton~202, 3C323.1, 4C09.72)}

\maketitle

\begin{abstract}
Optical spectropolarimetry carried out by \citet{kis04} has shown that several luminous type~1 quasars show a strong decrease of the polarized continuum flux in the rest-frame near-ultraviolet wavelengths of $\lambda<4000$\AA.
In the literature, this spectral feature is interpreted as evidence of the broadened hydrogen Balmer absorption edge imprinted in the accretion disk thermal emission due to the disk atmospheric opacity effect.
On the other hand, the quasar flux variability studies have shown that the variable continuum component in UV-optical spectra of quasars, which
is considered to be a good indicator of the intrinsic spectral shape of the accretion disk emission, generally have significantly flat spectral shape throughout the near-ultraviolet to optical spectral range.
To examine whether the disk continuum spectral shapes revealed as the polarized flux and as the variable component spectra are consistent with each other, we carry out multi-band photometric monitoring observations for a sample of four polarization-decreasing quasars of \citet{kis04} (4C09.72, 3C323.1, Ton~202, and B2~1208+32) to derive the variable component spectra and compare the spectral shape of them with that of the polarized flux spectra.
Contrary to expectation, we confirm that the two spectral components of these quasars have totally different spectral shape in that the variable component spectra are significantly bluer compared to the polarized flux spectra.
This discrepancy in the spectral shape may imply either (1) the decrease of polarization degree in the rest-frame ultraviolet wavelengths is not indicating the Balmer absorption edge feature but is induced by some unknown (de)polarization mechanisms, or (2) the ultraviolet-optical flux variability is occurring preferentially at the hot inner radii of the accretion disk and thus the variable component spectra do not reflect the whole accretion disk emission.
\end{abstract}

\section{Introduction}

\subsection{Optical polarization in luminous type~1 quasars}

Although there is no doubt that Active Galactic Nuclei (AGNs) play a key role in galaxy evolution across the cosmic time, the basic physics of the AGN central engine, namely the accretion disk around the supermassive black hole, and the structures in its vicinity have not yet been well understood \citep{ant13,ant15}.
One crucial difficulty in studying AGN accretion disks is that the Big Blue Bump (BBB) of the AGN accretion disk UV-optical continua is usually hidden under the so called ``small blue bump'' made up of Balmer continuum and Fe\emissiontype{II} lines from broad line region (BLR) (e.g., \cite{gra82}), and thus it is essentially impossible to quantify intrinsic AGN accretion disk continuum emission (e.g., \cite{kis08}).

In this context, optical spectropolarimetry offers an unique way to examine accretion disk emission in quasars.
It is well known that type 2 AGNs/quasars show strong polarization perpendicular to the radio structure due to the scattered AGN lights from the polar-scattering region (e.g., \cite{ant85}).
On the other hand, for type 1 quasars, weak linear polarization parallel to the radio structure is usually observed (e.g., \cite{sto79,ant83,sto84,sch00,kis04}), which cannot be explained by the polar-scattering or the polarization induced within the atmosphere of plane-parallel scattering-dominated disk (e.g., \cite{ant88,kis03})\footnote{The intrinsic AGNs/quasar accretion disk emission is claimed by some authors to be unpolarized because of the strong Faraday depolarization with magnetic fields in the disk atmosphere (e.g., \cite{ago96,sil09}, and references therein).}.
Currently, observed properties of polarization in type 1 quasars are understood such that the polarized flux is the electron-scattered accretion disk emission from the geometrically- and optically-thin equatorial scattering region located inside the dust torus (e.g., \cite{sto79,ant88,smi04,smi05,goo07,bat11,gas12,mar13,hut15}), although the synchrotron emission from the relativistic jet core viewed at larger angles to the ejection axis may explain the optical polarization in some type 1 quasars \citep{sch00}.

\begin{figure}[tbp]
\center{
\includegraphics[clip, width=3.2in]{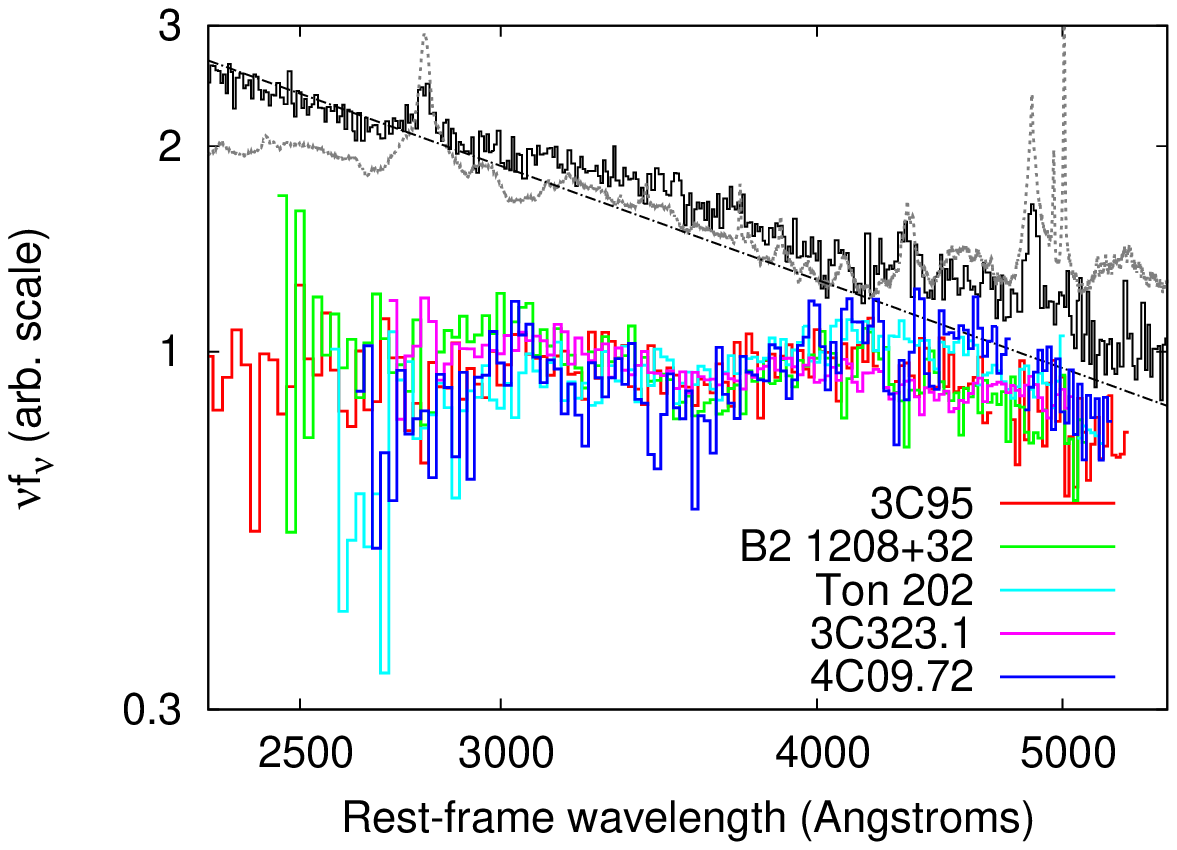}
}
 \caption{Polarized flux spectra for five quasars showing polarization decrease at $\lambda<4000$\AA\ (Figure~35 of \cite{kis04}). For comparison, the composite difference spectrum (i.e., the composite variable component spectrum) generated by using two epoch spectra of the SDSS quasars \citep{rua14} (the black thin solid line), and the composite spectrum of the SDSS quasars \citep{van01} (the gray dashed line) are also shown. The dash-dot line indicates a power-law spectrum with $\alpha_{\nu}=+1/3$ (in the unit of $f_{\nu}\propto \nu^{\alpha_{\nu}}$).}
 \label{fig:composite_specs}
\end{figure}

In many type~1 Seyfert galaxies and low luminosity quasars, the broad lines are also polarized but often at lower polarization degree and at different position angle than continuum, showing polarization angle rotation as a function of wavelength (e.g., \cite{smi04})
This implies that the equatorial scattering region is smaller or similar in size to the BLR (e.g., \cite{ang80,smi05,kis08b,bal16}).
By extending this interpretation of the scattering geometry, it is naturally expected that, as the BLR locates at more and more outer radius in higher luminosity quasars (e.g., \cite{ben13}), the BLR emission lines are to be depolarized in higher luminosity quasars in which the BLR locates more and more outer radius than the equatorial scattering region (see e.g., \cite{kis04,kis08b}).
Actually, \citet{kis03} and \citet{kis04} have carried out deep spectropolarimetry for 16 luminous quasars using Keck/LRIS and VLT/FORS1, and confirmed that the polarization of five of these quasars (3C95, Ton~202, B2~1208+32, 3C323.1, and 4C09.72) were confined to the continua, i.e., the BLR emission is depolarized (Figure~\ref{fig:composite_specs}) (see also  \cite{ant88,sch00}).
It should also be noted that the continuum polarization position angles of these five quasars are approximately wavelength independent, suggesting that the observed polarization is attributable to a single polarization source.
Furthermore, \citet{kis04} discovered that the polarized flux spectra of these five quasars were showing a decrease in the UV wavelength range of $\lambda<4000$\AA\ (see Figure~\ref{fig:composite_specs} and Table~\ref{obstarget}), which they interpreted as evidence of the broadened hydrogen Balmer absorption edge imprinted in the accretion disk thermal emission due to the disk atmospheric opacity effect (see also \cite{gas09,hu12}).
The presence of the hydrogen Balmer absorption edge in the AGN accretion disk emission has been naturally predicted by non-LTE radiative-transfer thermal accretion disk models of, e.g., \citet{hub00}, but the observational examination of it is very difficult due to the strong flux contamination from the UV Fe\emissiontype{II} pseudo-continuum and Balmer continuum emission from the BLR at $\lambda=$2200\AA-4000\AA. 
\citet{kis04} argued that the quasar spectropolarimetry has a potential to reveal the intrinsic spectral shape of the accretion disk thermal emission for each individual luminous type~1 quasar because the contaminating BLR emission does not appear in the polarized flux spectrum and thus the wavelength-independent electron (Thomson) scattering produces a scaled copy of the intrinsic accretion disk spectrum as the polarized flux component.
In subsequent works \citep{kis05,kis08}, they further confirmed that the measurements of the near-infrared (NIR) polarized fluxes of the UV polarization-decreasing quasars were consistent with those expected from the power-law extrapolation of the optical polarized flux spectra to the NIR region with the power-law index of $\alpha_{\nu}=1/3$, as predicted by the standard picture of the thermal optically-thick accretion disk models \citep{lyn69,pri72,sha73,nov73,hub00}.
Although it is currently unclear how common the continuum-confined polarization and its decrease at $\lambda<4000$\AA\ are for the population of luminous type~1 quasars, these results from the quasar spectropolarimetry, if confirmed, have profound implications for general understanding of the AGN accretion disk physics.

\subsection{Optical variability in luminous type~1 quasars}
\label{intro_variability}

Other than the polarimetry, optical flux variability, which is an ubiquitous property of AGNs, is also worthy of investigation in connection with the intrinsic spectral shape of the AGN accretion disk spectra.
Since the variability amplitude is enormous (typically $\sim$ 10-20\%), the variable component in the AGN UV-optical spectra must be reflecting the main energy source of AGN, namely the accretion disk emission itself \citep{gas08}.
Moreover, the strong inter-band correlations of the AGN variability indicate that the AGN UV-optical variability is not caused by localized independent fluctuations (having their own temperatures), but is a consequence of some kind of global changes in the accretion disk \citep{kok15}.
Therefore, it is natural to expect that the spectral shape of AGN accretion disk continuum can be obtained directly as the variable continuum spectral component (\cite{per06,li08,sch12,kok14,rua14}, and references therein).
It should be noted that the BLR emission is less variable than the underlying continuum, known as the intrinsic Baldwin effect \citep{kin90}; \citet{wil05} showed that the BLR emission lines vary at most 20\%-30\% as much the continuum emission, and \citet{kok14} noted that the low ionization lines of Fe\emissiontype{II} and Mg\emissiontype{II} are even less variable than the high ionization lines.
Since the flux contribution from the host galaxy and most of the BLR emission is non-variable, the spectral shape of the variable component spectra can be derived without suffering from the heavy spectral distortion by these contaminants.

In the NIR wavelength range, several spectral variability studies suggest that the accretion disk spectra revealed as the variable spectral component has spectral shape as blue as the thermal accretion disk model prediction of $\alpha_{\nu}=1/3$ (\cite{tom06,lir11,kos14}, and references therein).
In the near-ultraviolet to optical wavelength range, thermal accretion disk models generally predict the spectral turnover to much redder spectral slope (e.g., \cite{kis08}).
However, \citet{kok14} examined the spectral variability of $\sim$9000 Sloan Digital Sky Survey (SDSS) quasars using $\sim$10 years multi-band light curves available in SDSS Stripe~82 region, and found that the variable component spectra of quasars are well described by the power-law shape with $\alpha_{\nu}\sim 1/3$ even in the UV-optical wavelength range, and thus are systematically too blue to be explained by the existing thermal accretion disk models even after considering in the flux contamination from the (weakly) variable broad emission lines, Fe\emissiontype{II} emission lines, and the Balmer continuum emission (see also \cite{sch12,rua14}, and Figure~\ref{fig:composite_specs}).
It should also be noted that the strong dip feature at $\lambda<4000$\AA\ observed in the polarized flux seems to be absent in the composite variable continuum spectrum of SDSS quasars \citep{wil05,per06,kok14,rua14}.
In summary, quasar variability studies suggest that the intrinsic accretion disk spectrum revealed as the variable continuum component cannot be explained by existing accretion disk models.

\subsection{The goal of this study}
\label{intro_goal}

As described above, the results from the two kinds of studies seem to be contradicting:
on the one hand polarimetric studies suggest that the intrinsic accretion disk spectrum revealed by the spectropolarimetry has a spectral shape consistent with the thermal optically-thick accretion disk model predictions of , e.g., \citet{hub00}, but on the other hand variability studies suggest that the intrinsic accretion disk spectrum obtained as a variable continuum component has too blue ultraviolet spectral shape to be explained by existing accretion disk models.
This discrepancy, if confirmed, is a huge problem in that the basic assumptions underlying these observations are related to our fundamental understandings of quasar central engine, and thus definitely offers new insights into the nature of the quasar/AGN BBB emission.

However, we should note that the above statement is currently based on the observations for different quasar samples.
Although the quasars in the \citet{kis04}'s sample have similar optical spectra with the SDSS quasars (actually B2~1208+32, Ton~202, and 3C323.1 are contained in the spectroscopically-confirmed SDSS quasar catalog; \cite{sch10}), on the one hand the variability studies are based on the general population of several thousands of SDSS quasars, but on the other hand the spectropolarimetric studies are based on the small, and probably biased, sample of quasars (see \cite{sch00}; this point will be discussed elsewhere).
The purpose of this work was to examine the optical spectral variability for the quasars showing the Balmer edge-like feature in their polarized flux spectra confirmed by \citet{kis04}, and probe whether or not these quasars also show flat variable component spectra as the other normal quasars or show.
Here we present the multi-band optical light curves for four of the five quasars in \citet{kis04}'s sample (4C09.72, B2~1208+32, Ton~202, and 3C323.1), and compare the spectral shape of the polarized flux and the variable flux components of them.

This paper is organized as follows.
In Section~\ref{data}, we present multi-band optical light curves of the four quasars obtained by using the 1.05-meter Kiso Schmidt telescope at the Kiso Observatory in Japan.
Then, we derive the variable component spectra for the quasars from the multi-band light curves by using ``flux gradient method'' in Section~\ref{analysis}.
In Section~\ref{discussion}, we compare the spectral shape of the the variable and polarized component spectra of these quasars in detail (Section~\ref{note1}), and discuss about the possible interpretations of the relationship between these two spectral components (Section~\ref{note2}).
Finally, summary and conclusions are given in Section~\ref{summary}.

\section{Multi-band photometric observations with the 1.05-meter Kiso Schmidt telescope}
\label{data}

\begin{table*}
\tbl{List of our target quasars with polarization dip in $\lambda<4000$\AA.}{%
\begin{tabular}{lccccc}  
\hline\noalign{\vskip3pt} 
\multicolumn{1}{c}{Name} & Coordinate & $E(B-V)$ & $z$ & $P$ (per cent) at 2891-3600\AA & $P$ (per cent) at 4000-4731\AA \\  [2pt] 
\hline\noalign{\vskip3pt} 
B2~1208+32 & 12:10:37.56 +31:57:06.02 & 0.017 & 0.388 & 1.01 $\pm$ 0.01& 1.41 $\pm$ 0.01\\
Ton~202    & 14:27:35.60 +26:32:14.55 & 0.019 & 0.366 & 1.25 $\pm$ 0.01& 1.87 $\pm$ 0.02\\
3C323.1    & 15:47:43.53 +20:52:16.66 & 0.042 & 0.264 & 1.51 $\pm$ 0.01& 2.19 $\pm$ 0.01\\
4C09.72    & 23:11:17.74 +10:08:15.77 & 0.042 & 0.433 & 0.50 $\pm$ 0.01& 0.93 $\pm$ 0.01\\  [2pt] 
\hline\noalign{\vskip3pt} 
\end{tabular}}\label{obstarget}
\begin{tabnote}
\hangindent6pt\noindent
\hbox to6pt{\footnotemark[$*$]\hss}\unskip% 
 The redshift $z$ and polarization degree $P$ (per cent) at two rest-frame wavelength ranges are taken from Table~6 of \citet{kis04}. The reported values of $P$ for 4C09.72 and 3C323.1 are corrected for the inter stellar polarization effect. The Galactic selective extinction $E(B-V)$ based on \citet{sch98} is taken from NASA's Extragalactic Database (NED).
\end{tabnote}
\end{table*}

We carried out $u$, $g$, $r$, $i$, and $z$-bands photometric monitoring observations for four of the five quasars with polarization dip in $\lambda<4000$\AA\ in \citet{kis04}'s sample (4C09.72, B2~1208+32, Ton~202, and 3C323.1; see Table~\ref{obstarget}) from April 2015 to February 2016, using the 1.05-meter Kiso Schmidt telescope being operated by the Kiso Observatory of the University of Tokyo.
The 1.05-meter Kiso Schmidt telescope is equipped with Kiso Wide Field Camera (KWFC; \cite{sak12}), which has eight CCD chips (the field of view is 2.2 degree $\times$ 2.2 degree) composed of four CCDs with 2k $\times$ 4k pixels manufactured by Lincoln Laboratory, Massachusetts Institute of Technology (MIT) and four ST-002A CCDs with 2k $\times$ 4k pixels manufactured by Scientific Imaging Technologies, Inc. (SITe).
The average wavelengths of the Kiso/KWFC photometric bands are not precisely known, thus we assume them to be identical with the SDSS imaging camera, i.e., 
3551, 4686, 6166, 7480, and 8932\AA\ for $u$, $g$, $r$, $i$, and $z$-bands, respectively, taken from the SDSS web site\footnote{http://www.sdss.org/instruments/camera/} (see also \cite{fuk96,doi10}).
The four target quasars are located within the SDSS Legacy survey footprint \citep{yor00,gun06} and thus the photometric calibration can be done by using the SDSS photometry data of the field stars on the same frames with the targets\footnote{Although we have also been observing the other quasar 3C95 since September 2015, we do not include it in the sample of this work because it is not in the SDSS Legacy survey footprint thus currently the absolute flux calibration cannot be performed. Further observations on this object and the construction of the reference star catalog around it are in progress and will be reported elsewhere.}.

Our observations have been carried our in queue mode.
In the queue mode, calibration data, i.e., dome flat frames for every filter and bias frames, are usually obtained automatically at the beginning and/or the end of each observing night.
Since the dark current of the KWFC CCDs is below 5 $e^{-} [\rm{hour}^{-1} \rm{pixel}^{-1}]$ at an operation temperature of 168 K (e.g., \cite{sak12}), we did not obtain dark frames.
The five band images were normally obtained quasi-simultaneously in $giruz$ order with four dithering pointing for each band, although the number of the bands and exposures was reduced in several observing nights according to the weather condition.
The exposure time for each image was 30 sec for every filter in April-June 2015, and is 60 sec for $g$, $r$, and $i$-band and 120 sec for $u$ and $z$-band in July 2015-February 2016.

For the observations presented in this paper, we used 2$\times$2 binning four MIT chips FAST readout mode (1.88 arcsec/pixel) to reduce the overhead time, and the targets were acquired on to the MIT chip\#3 (named as MIT-4 in Figure~2 of \cite{sak12}).
Thus, in this work we only reduce and analyse the chip\#3 images.
The detector temperature is kept to be 167.9-168.0 K during the observations.
Each of the 2k$\times$4k CCDs of KWFC has a dual amplifier readout.
During the data reduction, we treat the two readout areas on the chip\#3 (upper 1k$\times$4k and lower 1k$\times$4k pixels, corresponding the two amplifiers) separately.
During the data reduction, the gain factor and the readout noise of the chip\#3 is assumed to be 2.3 electron/ADU and 15 electrons, respectively, for both of the readout ports. 
Overscan subtraction for all of the calibration and object frames is carried out by using the column overscan region with the use of IRAF task {\tt colbias}\footnote{IRAF is distributed by the National Optical Astronomy Observatory, which is operated by the Association of Universities for Research in Astronomy (AURA) under a cooperative agreement with the National Science Foundation.}.
The master bias frame for each of the observing night is generated by median-combining the overscan-subtracted bias frames taken at the same or the closest night, and is used to correct for the large scale bias pattern.
Master dome-flat frames for each of the observing filters are created by median-combining the dome-flat frames taken at the same or the closest night by normalizing the large scale sensitivity and/or illumination pattern with the use of the IRAF task {\tt mkillumflat}.
These master dome-flat frames are used to correct for the pixel-to-pixel sensitivity variation of the CCD detector.
To correct for the global sensitivity inhomogeneity of the detector, we use super sky-flat frames created by median-combining different nights' object images by masking detected objects by the use of the IRAF task {\tt objmasks}.
During the creation of the master dome-flat and super sky-flat calibration frames, the dead pixel lines at the right bottom portion of the chip\#3 ($X=1263$-$2073$ and $Y=33$-$41$ pixel coordinates in the 2$\times$2 binning mode) are replaced by linear interpolation along columns by using the IRAF task {\tt fixpix}.

Sky background subtraction, source extraction and aperture photometry for the overscan, bias, and pixel-to-pixel and global flat-fielded object frames were carried out by using SExtractor (version 2.8.6; \cite{ber96}), and coordinate conversion (shifts, pixel scale, and rotation) was carried out by using WCSTools imwcs (version 3.8.4; \cite{min06}) and USNO-B1.0 catalog \citep{mon03}.
Automatically interpolated background-map is calculated and subtracted from the object frames (namely, we use the SExtractor parameters of BACK\_TYPE=AUTO, BACKPHOTO\_TYPE=GLOBAL, BACK\_SIZE=64, and BACK\_FILTERSIZE=6).
The dead pixel lines mentioned above are masked during the SExtractor runs.
Considering the seeing statistics at the Kiso site (3.9 arcsec FWHM at median in $g$-band; \cite{morokuma14}), the extraction aperture is set to 5 pixels ($\sim$ 9.4 arcsec).
The detection and analysis thresholds are set to be 3$\sigma$, where $\sigma$ is defined as the sum of the photon noise and the readout noise, not including noises introduced by the overscan/bias subtraction and the flat-fielding processes.
It should be noted that the target quasars are observed as point-like objects, thus the apparent flux changes due to seeing variations are cancelled by the relative photometry using field stars described below.

The magnitude zero-point shift for each frame was evaluated by taking a 3$\sigma$-clipping weighted average of the differences between the instrumental magnitudes and the SDSS model magnitudes (retrieved from the SDSS Data Release 12 (DR12) SkyServer Star view; \cite{ala15}) of field stars within the field of view of each of the readout area on the chip\#3 (1.076 degrees $\times$ 0.269 degrees).
The field stars flagged as CLEAN in the SDSS database were used, and the object frames with no more than 30 position-matched field stars were excluded from the analysis.
It is known that the photometric zeropoints of the SDSS magnitude system are slightly offset from those of the AB magnitude system, thus we apply the recommended offset values of $u_{AB} = u_{SDSS} - 0.04$ mag, and $z_{AB} = z_{SDSS} + 0.02$ mag to our calibrated data, following description in the SDSS website \footnote{http://www.sdss.org/dr12/algorithms/fluxcal/}.
The magnitude to flux conversion is done by assuming the zero magnitude flux as 3631 Jansky (Jy).
Since the color-term correction factors for the KWFC SDSS filters have not been determined, in this work we do not apply the color correction. 
We also decide not to combine the (mostly single-epoch) SDSS Legacy Survey photometric data for the target quasars with our light curves to avoid the possible systematic error due to the non-correction of the color-term.

\begin{figure}[tbp]
\center{
\includegraphics[clip, width=3.2in]{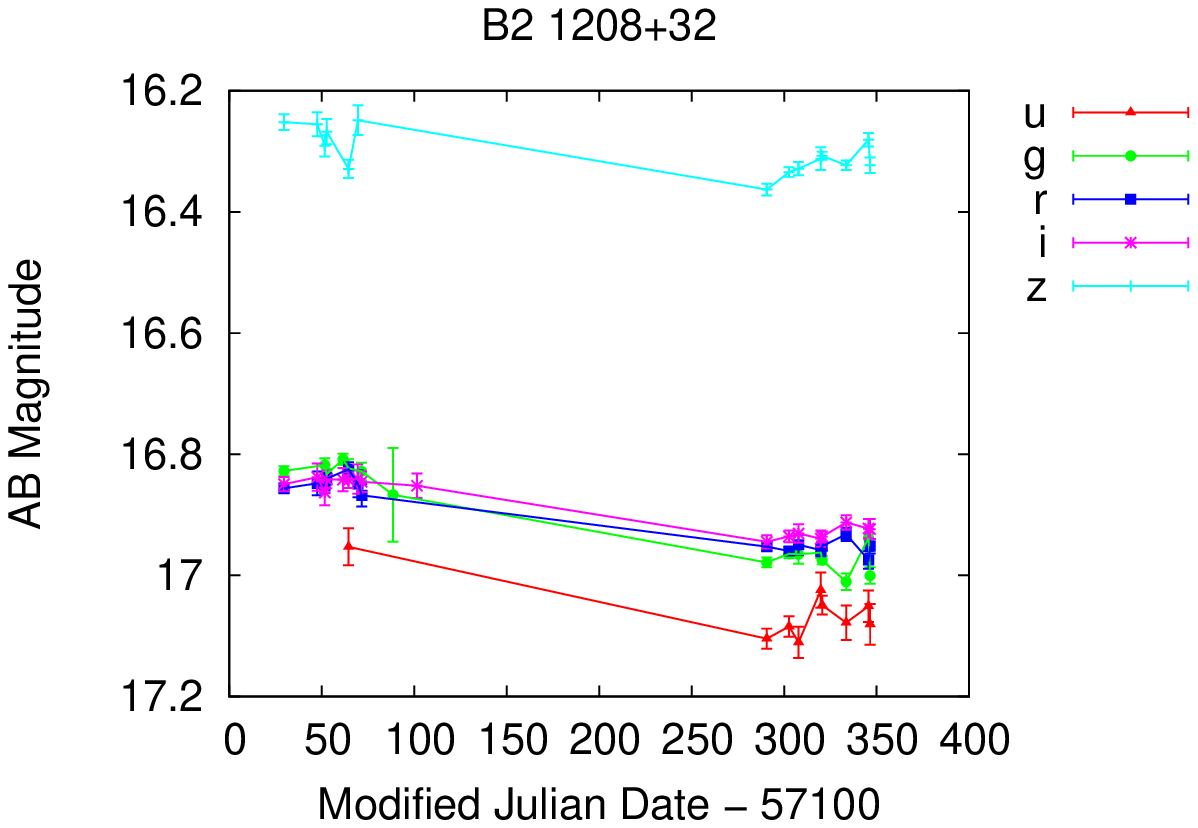}
\includegraphics[clip, width=3.2in]{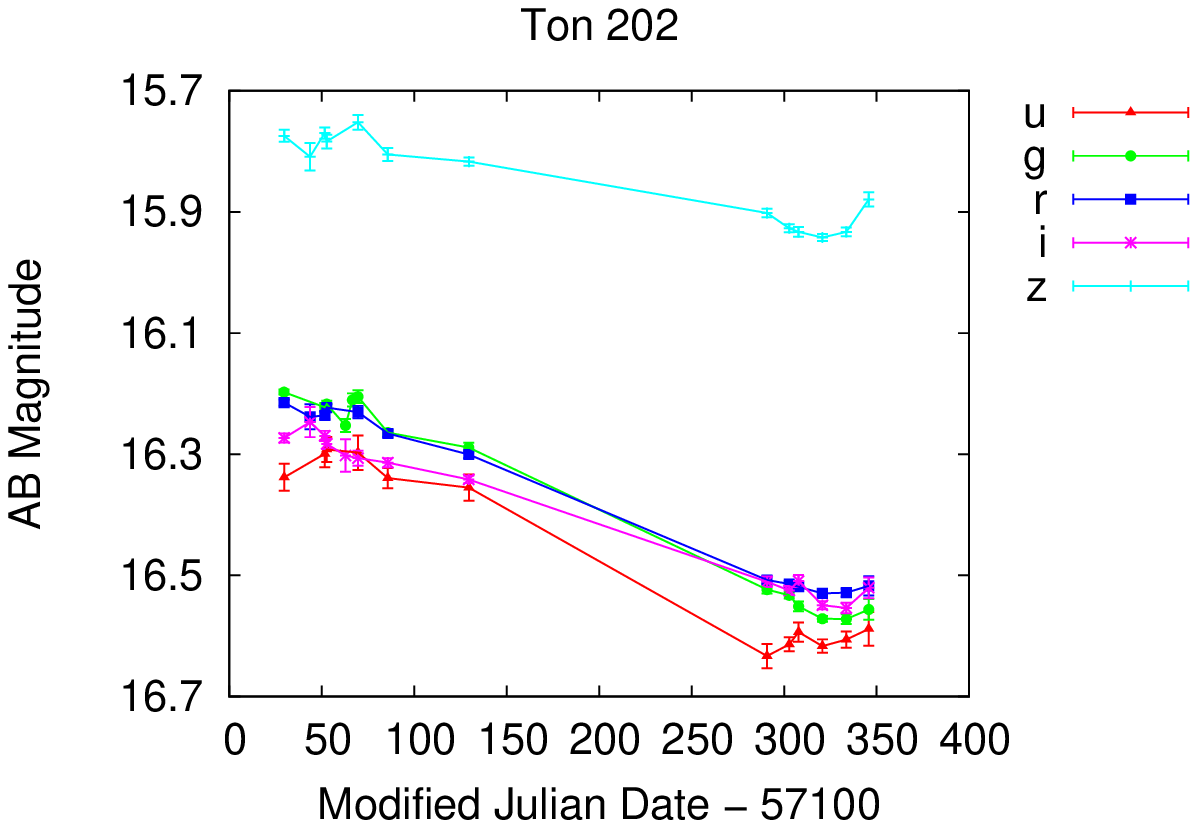}
\includegraphics[clip, width=3.2in]{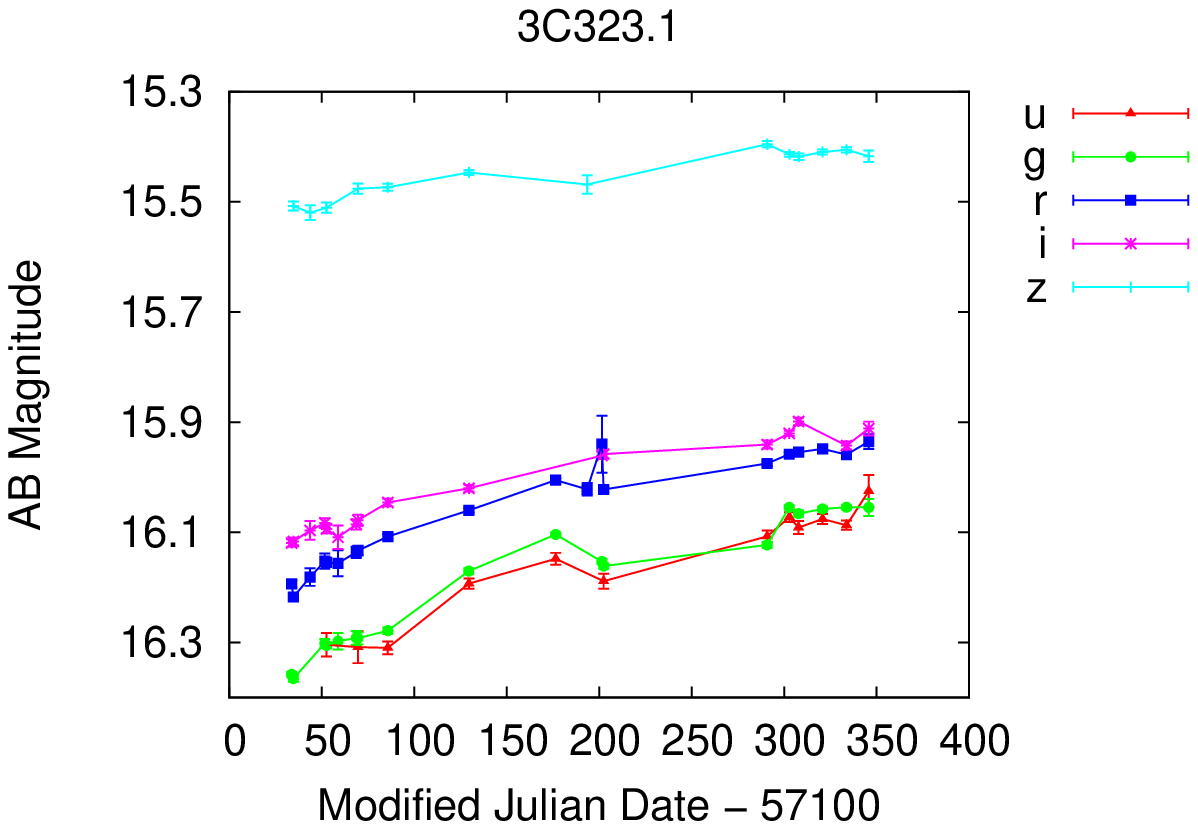}
\includegraphics[clip, width=3.2in]{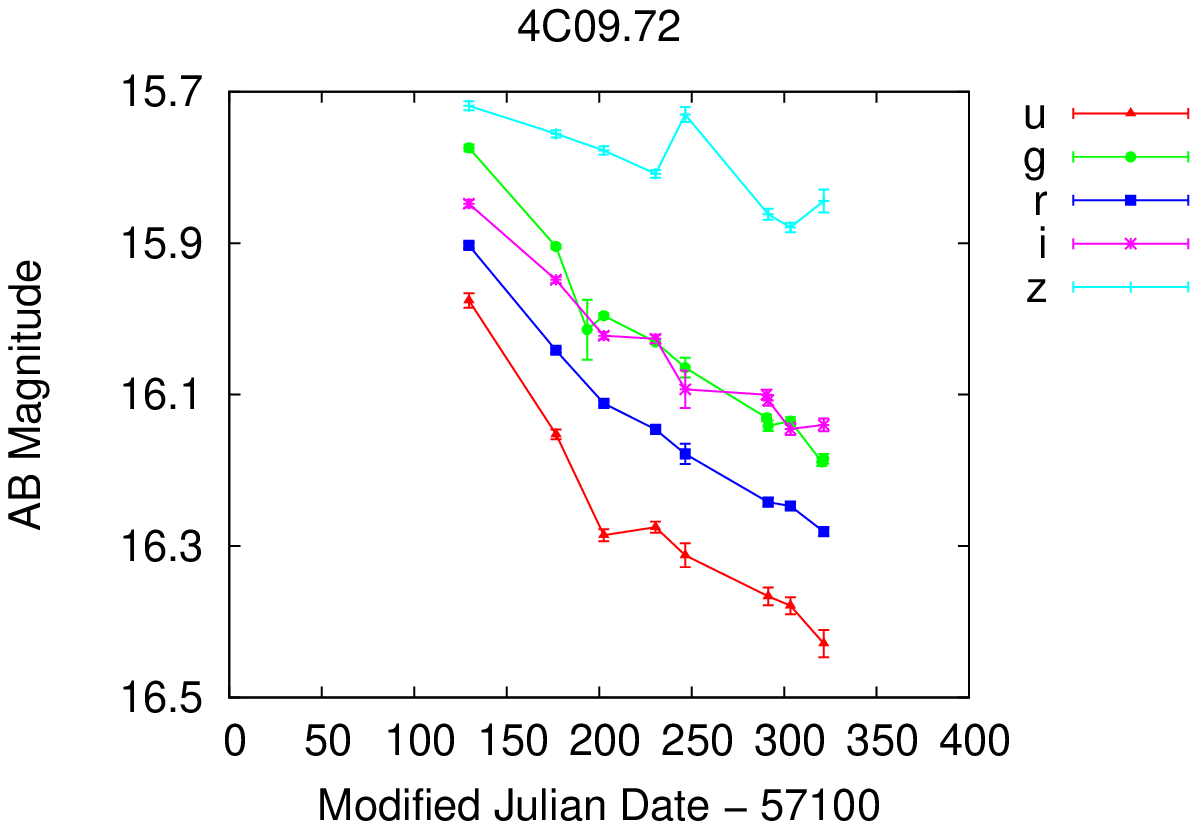}
}
 \caption{$u$, $g$, $r$, $i$, and $z$-band light curves of B2~1208+32, Ton~202, 3C323.1, and 4C09.72, obtained at the Kiso observatory. Galactic extinction is not corrected.}
 \label{fig:mag_lightcurve}
\end{figure}

The obtained light curves are shown in Table~\ref{lightucrve_table} and plotted in Figure~\ref{fig:mag_lightcurve}.
The reported magnitude/flux values for a given filter are the weighted averaged values of the dithering images taken at the same nights.
The observation epoch for each data point is expressed in the unit of Modified Julian Date (MJD) at the midpoint of the exposures.
We confirm that all of the four quasars show flux variability in all five bands during our observations, although the variability amplitude in B2~1208+32 is relatively small compared to the others.
The observed variability amplitudes of $\Delta g \sim $0.1-0.4 mag within $\sim$ 200 days in the quasar rest-frame are consistent with the general property of the SDSS quasars (e.g., \cite{mac12}).

\begin{table*}[tbp]
\tbl{KWFC $u$, $g$, $r$, $i$, and $z$-band light curves for B2~1208+32, Ton~202, 3C323.1, and 4C09.72.}{%
\begin{tabular}{lcccc}  
\hline\noalign{\vskip3pt} 
\multicolumn{1}{c}{Name} & MJD & Magnitude & Error in magnitude & Band \\  [2pt] 
\hline\noalign{\vskip3pt} 
B2~1208+32 & 57164.47822 & 16.9528 & 0.0306  &  $u$\\
B2~1208+32 & 57390.64625 & 17.1045 & 0.0167  &  $u$\\
$\cdots$   & $\cdots$    &$\cdots$ &$\cdots$ &  $\cdots$\\ [2pt]  
\hline\noalign{\vskip3pt} 
\end{tabular}}\label{lightucrve_table}
\begin{tabnote}
\hangindent6pt\noindent
\hbox to6pt{\footnotemark[$*$]\hss}\unskip% 
 The reported magnitude values are in AB magnitude system, and not corrected for Galactic extinction. A machine-readable version of the full table is available in the ``Other formats / Source'' in arXiv.
\end{tabnote}
\end{table*}

\section{Analysis}
\label{analysis}

\begin{figure}[tbp]
\center{
\includegraphics[clip, width=3.2in]{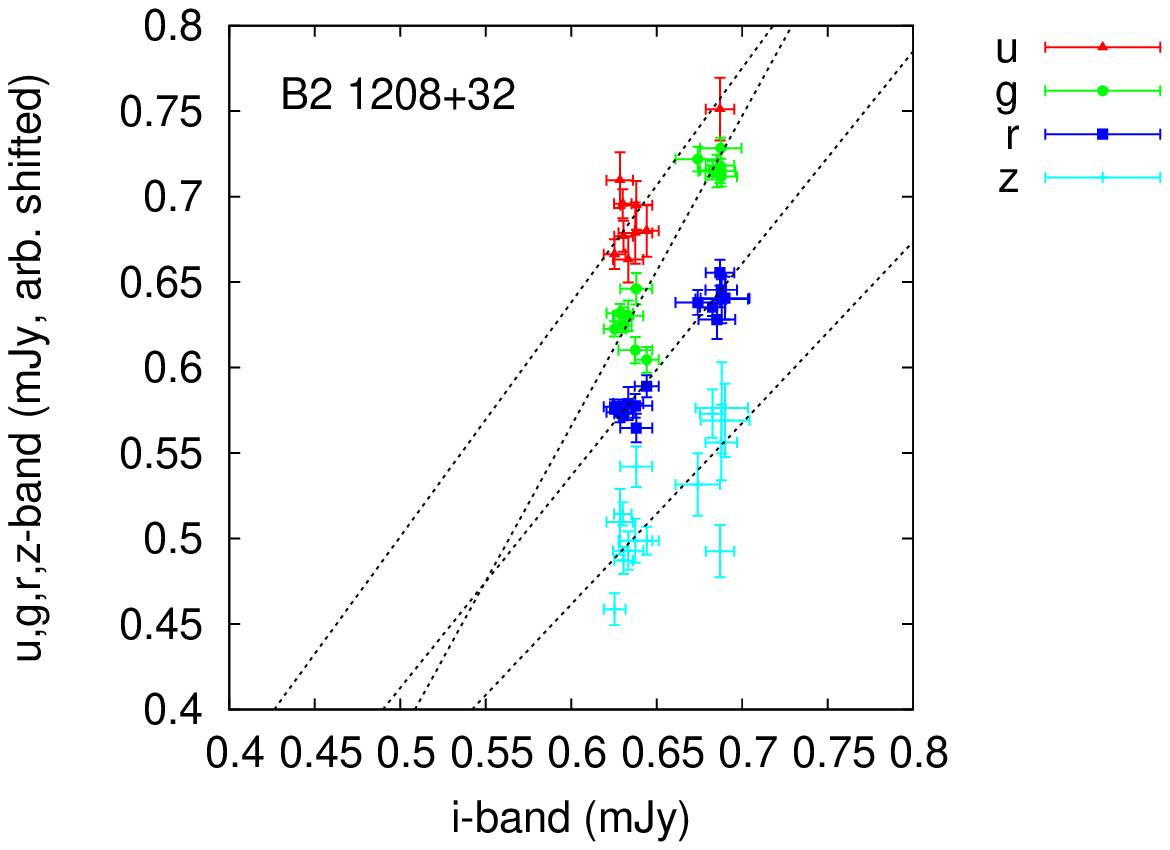}
\includegraphics[clip, width=3.2in]{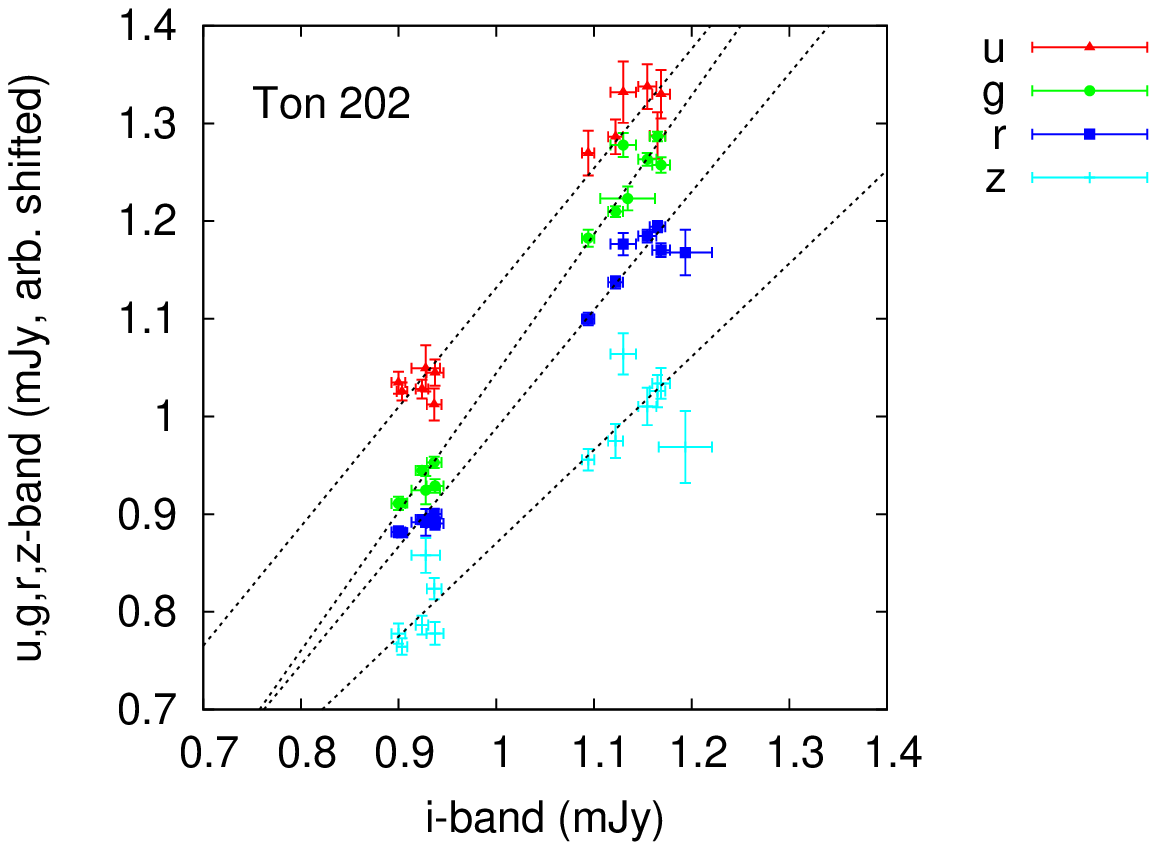}
\includegraphics[clip, width=3.2in]{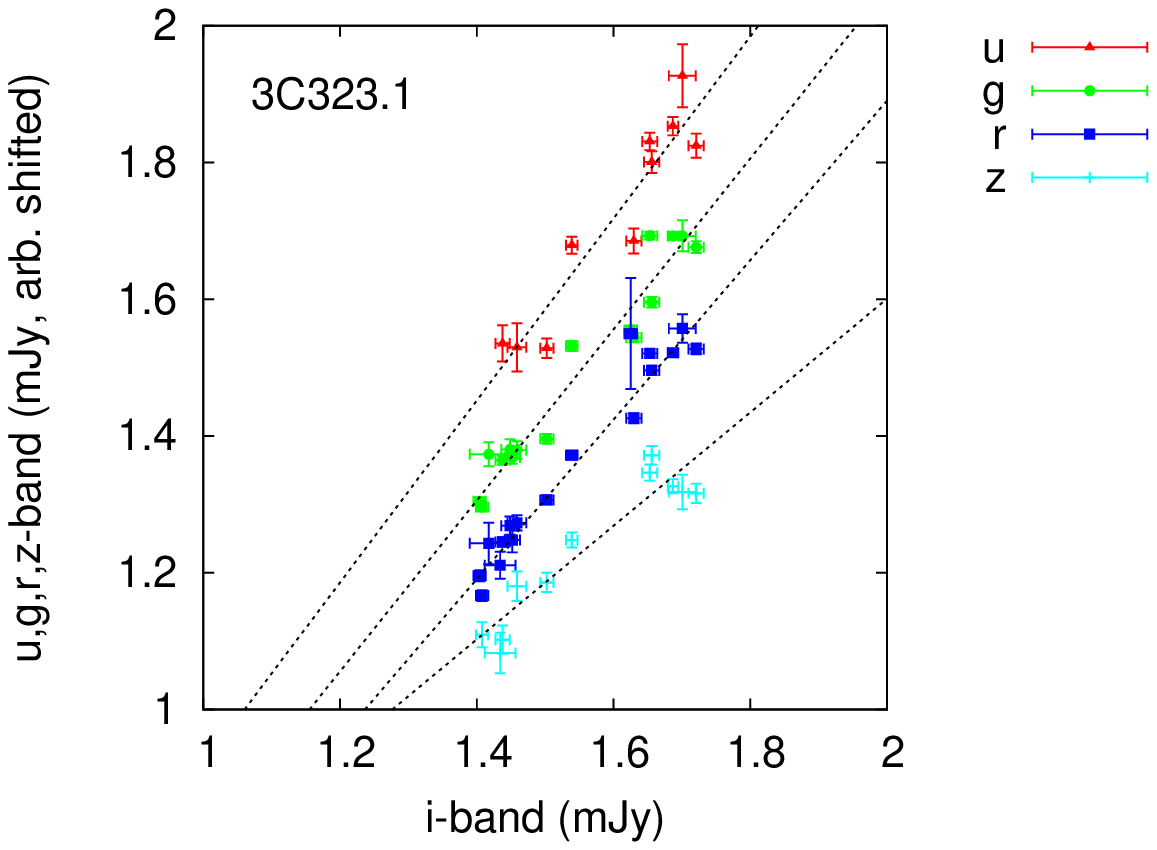}
\includegraphics[clip, width=3.2in]{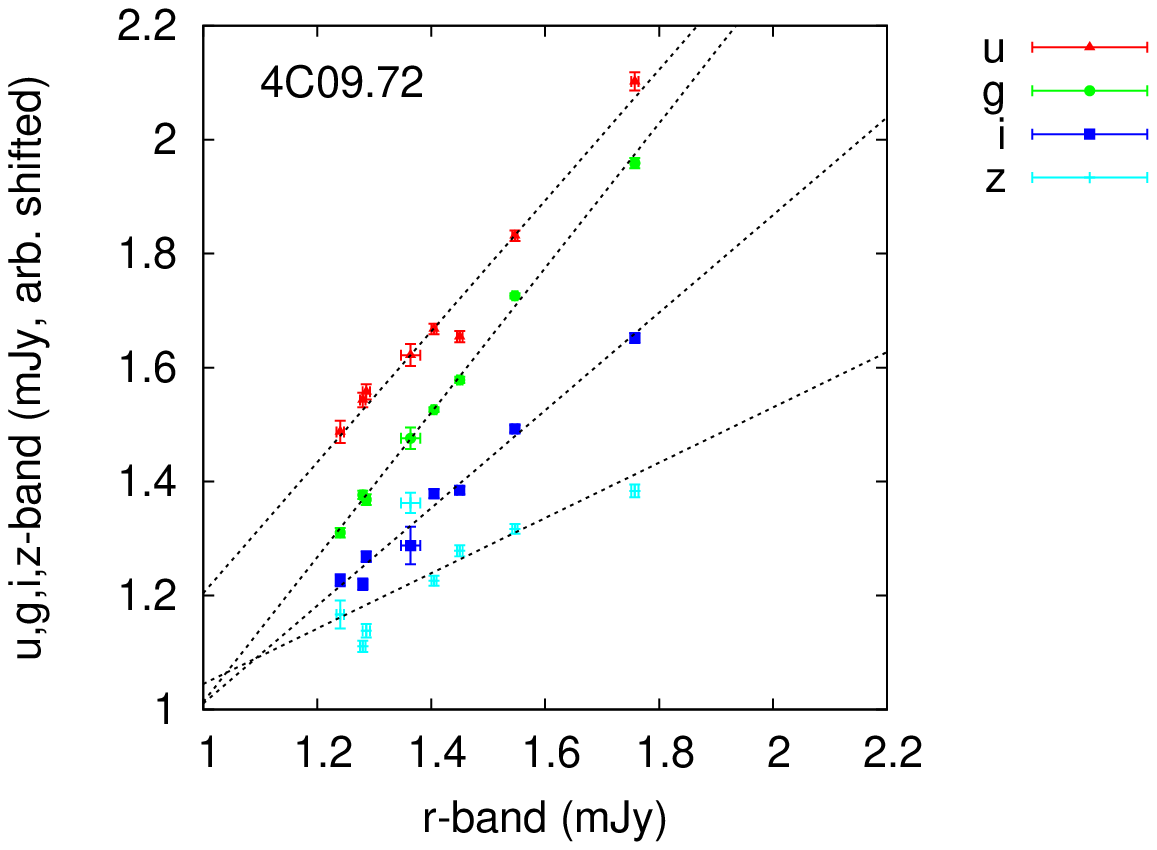}
}
 \caption{Galactic extinction-corrected flux-flux plots of the light curves of B2~1208+32, Ton~202, 3C323.1, and 4C09.72 obtained at the Kiso observatory. The flux values are arbitrarily shifted on the y-axis for clarity.}
 \label{fig:fluxflux}
\end{figure}

\begin{figure}[tbp]
\center{
\includegraphics[clip, width=2.8in]{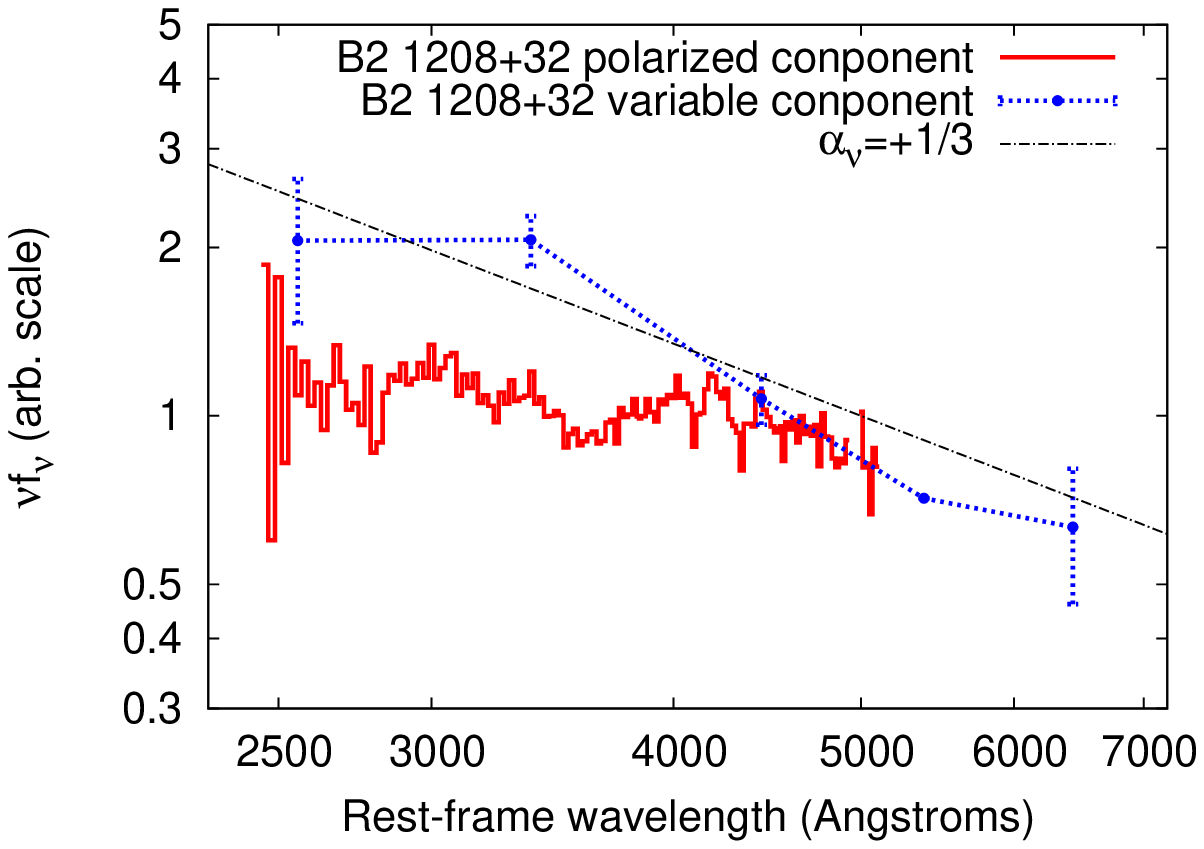}
\includegraphics[clip, width=2.8in]{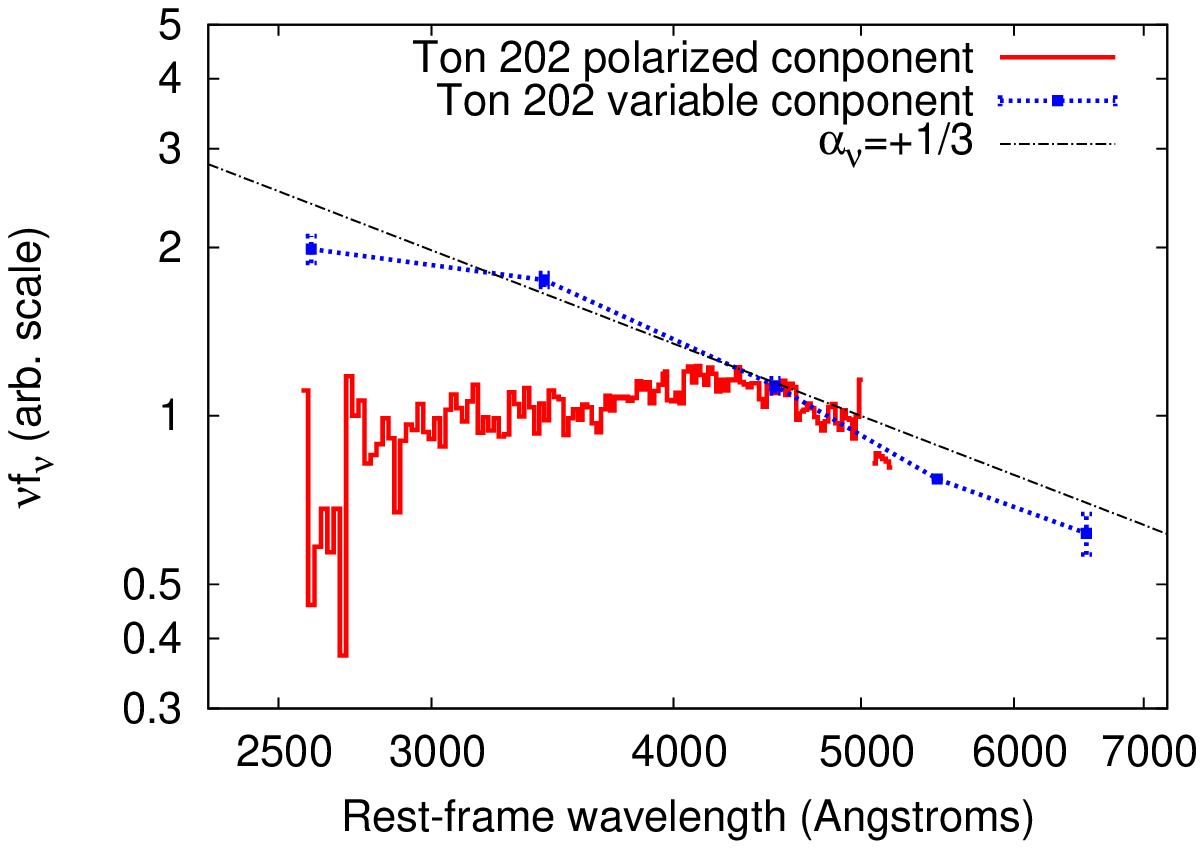}
\includegraphics[clip, width=2.8in]{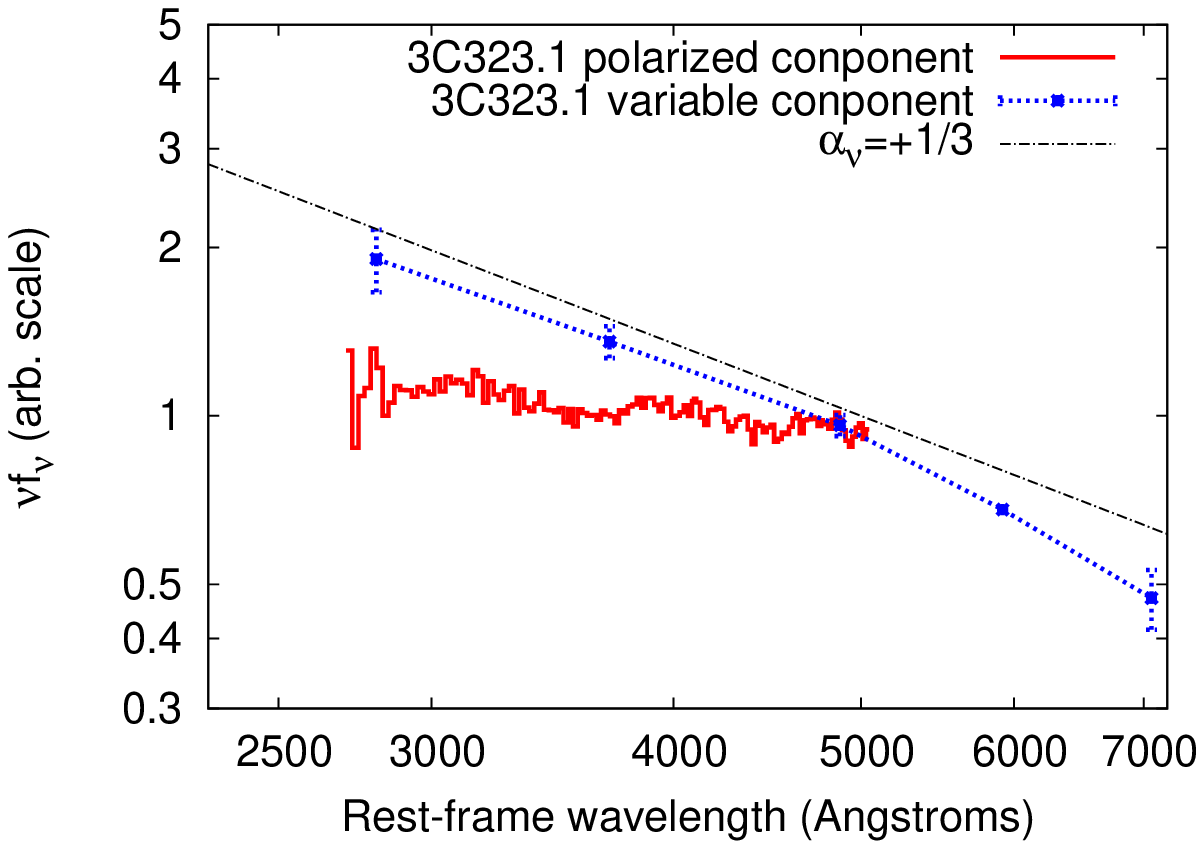}
\includegraphics[clip, width=2.8in]{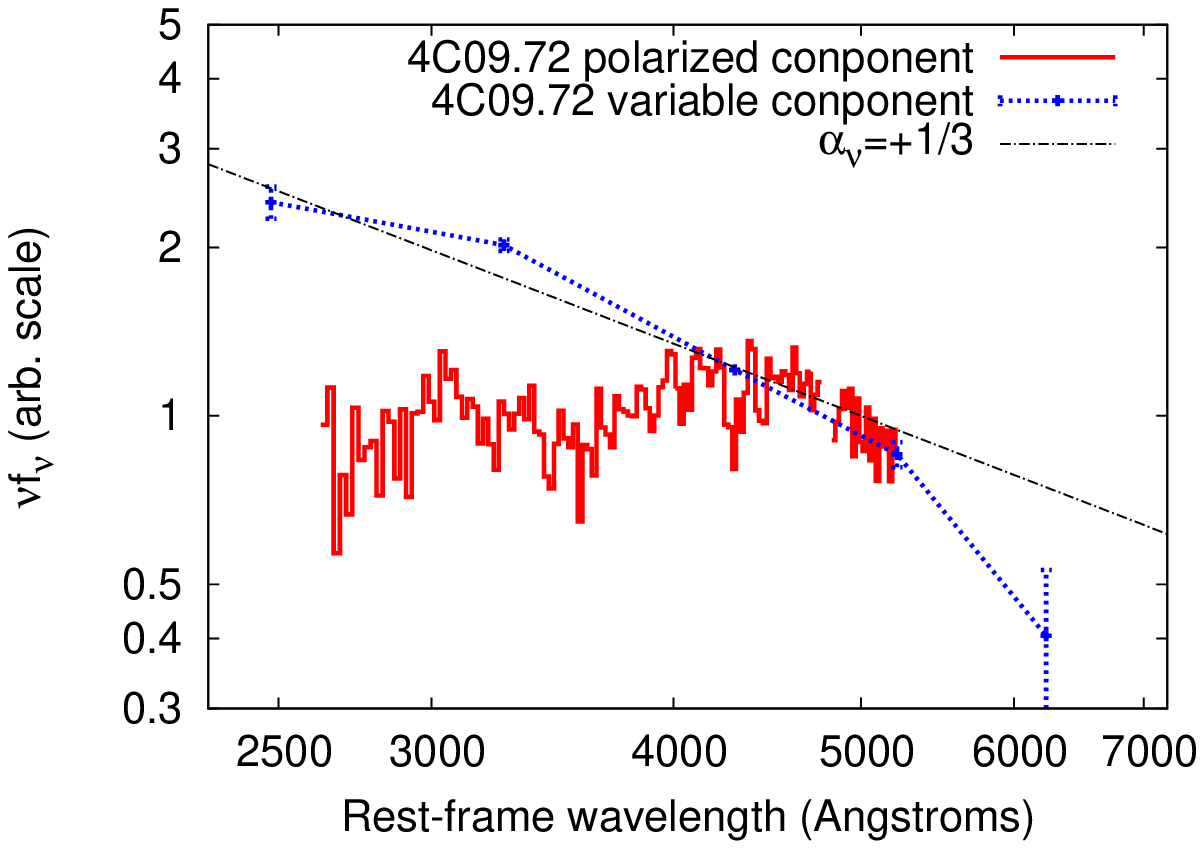}
}
 \caption{Comparison of the spectral shape of the variable and polarized components in the optical spectra of B2 1208+32, Ton 202, 3C323.1, and 4C09.72. Dotted lines indicate the variable component spectra derived in this work (see Section~\ref{analysis}), and the solid lines are the polarized component spectra taken from \citet{kis04}. The polarized component spectra are the same as Figure~\ref{fig:composite_specs}. 
The two spectra in each panel are scaled to approximately match each other at the red side of the polarized component spectrum.
For comparison, a power-law spectrum with $\alpha_{\nu}=+1/3$ (in the unit of $f_{\nu}\propto \nu^{\alpha_{\nu}}$) is also shown.}
 \label{fig:comparison_each}
\end{figure}

\begin{figure}[tbp]
\center{
\includegraphics[clip, width=3.2in]{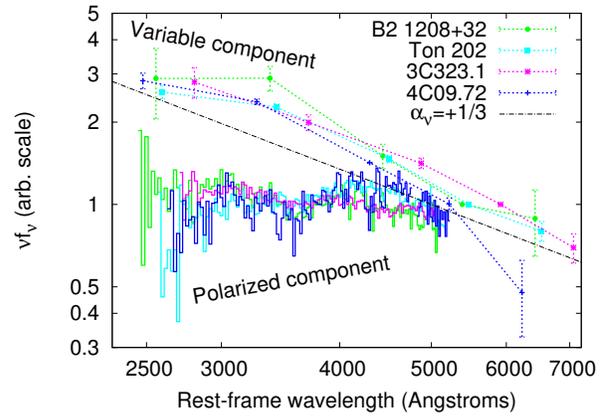}
}
 \caption{The same as in Figure~\ref{fig:comparison_each}, but all objects are shown simultaneously. The variable component spectra are scaled to 1 in $i$ band for clarity. The spectra of B2~1208+32, Ton~202, 3C323.1, and 4C09.72 are colored green, cyan, magenta, and blue, respectively.}
 \label{fig:comparison}
\end{figure}

\begin{table}
\tbl{The best-fit linear regression parameters (slope and intercept) for the two-band flux-flux plots.}{%
\begin{tabular}{lccc}  
\hline\noalign{\vskip3pt} 
\multicolumn{1}{c}{Name} & Band-pair ($x$-$y$) & Slope & Intercept [mJy] \\  [2pt] 
\hline\noalign{\vskip3pt} 
B2~1208+32  &   $i$-$u$  &  1.371 $\pm$ 0.395 &  $-$0.284 $\pm$ 0.251  \\
B2~1208+32  &   $i$-$g$  &  1.814 $\pm$ 0.186 &  $-$0.523 $\pm$ 0.121  \\
B2~1208+32  &   $i$-$r$  &  1.241 $\pm$ 0.126 &  $-$0.158 $\pm$ 0.081  \\
B2~1208+32  &   $i$-$z$  &  1.061 $\pm$ 0.288 &  $+$0.425 $\pm$ 0.187  \\\hline
Ton~202  &   $i$-$u$  &  1.222 $\pm$ 0.068 &  $-$0.220 $\pm$ 0.068  \\
Ton~202  &   $i$-$g$  &  1.421 $\pm$ 0.040 &  $-$0.376 $\pm$ 0.041  \\
Ton~202  &   $i$-$r$  &  1.210 $\pm$ 0.038 &  $-$0.172 $\pm$ 0.039  \\
Ton~202  &   $i$-$z$  &  0.955 $\pm$ 0.079 &  $+$0.716 $\pm$ 0.082  \\\hline
3C323.1  &   $i$-$u$  &  1.331 $\pm$ 0.169 &  $-$0.611 $\pm$ 0.272  \\
3C323.1  &   $i$-$g$  &  1.250 $\pm$ 0.082 &  $-$0.544 $\pm$ 0.127  \\
3C323.1  &   $i$-$r$  &  1.166 $\pm$ 0.051 &  $-$0.292 $\pm$ 0.079  \\
3C323.1  &   $i$-$z$  &  0.830 $\pm$ 0.102 &  $+$1.241 $\pm$ 0.160  \\\hline
4C09.72  &   $r$-$u$  &  1.148 $\pm$ 0.076 &  $-$0.244 $\pm$ 0.108  \\
4C09.72  &   $r$-$g$  &  1.269 $\pm$ 0.029 &  $-$0.156 $\pm$ 0.042  \\
4C09.72  &   $r$-$i$  &  0.857 $\pm$ 0.044 &  $+$0.304 $\pm$ 0.063  \\
4C09.72  &   $r$-$z$  &  0.485 $\pm$ 0.151 &  $+$1.160 $\pm$ 0.216  \\[2pt]
\hline\noalign{\vskip3pt} 
\end{tabular}}\label{regression_table}
\begin{tabnote}
\hangindent6pt\noindent
\hbox to6pt{\footnotemark[$*$]\hss}\unskip% 
The reported uncertainty is $\pm 1\sigma$.
\end{tabnote}
\end{table}

The flux-flux plots of two-band simultaneous light curves of AGNs and quasar in the UV-optical wavelength range are known to be well fitted by straight lines of $y = a + bx$, which indicates that the variable continuum component in AGNs and quasars keeps its spectral shape nearly constant over several years to several tens of years of the flux variability (\cite{sak10,sak11,kok14,kok15,ram15} and references therein, although see also \cite{sun14}).
This means that the linear regression slope (gradient) of the flux-flux plot of the quasar two-band light curves can be used as an indicator of the color of the variable component spectrum, since the flux gradient represents the flux ratio of the two-band fluxes of the variable component spectrum (``flux gradient method''; e.g., \cite{cho81,win92,hag97,win97,gla04,hag06,cac07,sak10,sak11,kok14,ram15}).
The good point of the flux gradient method is that the regression slope is not affected by the flux contamination from the non-variable spectral components (the host galaxy flux and time-averaged flux level of BLR emission).
Thus, by applying the flux gradient method to the multi-band light curves, we can derive the relative spectra of the variable component in quasars without suffering from the spectral distortion due to the flux contribution from the non-variable spectral components (see \cite{kok14}, and references therein).
Here we examine the spectral shape of the variable flux component in the four quasars by using the multi-band photometric monitoring data obtained at the Kiso observatory.
The two-band simultaneous light curve for each target and for each band-pair is generated by combining the two-band measurements taken at the same nights.

Figure~\ref{fig:fluxflux} shows the flux-flux plots of the light curves of B2 1208+32, Ton~202, 3C323.1, and 4C09.72, overplotted with the best-fit linear regression lines for each band-pair derived by using {\tt MPFITEXY} IDL routine \citep{wil10}.
The {\tt MPFITEXY} routine depends on the {\tt MPFIT} package \citep{mar09}, and is able to cope with the data with intrinsic scatter, which is automatically adjusted to ensure $\chi^2$/(degrees or freedom) $\sim 1$ (see \cite{tre02,nov06,par12} for details).
Flux values in Figure~\ref{fig:fluxflux} are corrected for the Galactic extinction (Galactic extinction values for the SDSS filters are taken from NED; \cite{sch98}).
For B2 1208+32, Ton202 and 3C323.1, we choose to use $i$-band flux as the reference of the continuum flux variability and as $x$-axis in the flux-flux plots (i.e., independent variable), since the $i$-band wavelength range is less contaminated by broad emission lines than the other bands.
For 4C09.72, whose observed-frame $i$-band wavelength range contains the H$\beta$ line, $r$-band flux is instead used as the reference of the continuum flux variability and as $x$-axis in the flux-flux plots.

The best-fit linear regression parameters (regression slope and intercept) for the flux-flux plots are summarized in Table~\ref{regression_table}.
Since the variability amplitude of B2~1208+32 is small and the sampling of the light curves is poor, the regression parameters are not well constrained for B2~1208+32, especially in the $i$-$u$ band-pair.
For the other quasars, the flux-flux plots are well fitted by the linear regression lines.
We should also note that the regression intercepts of $i$-$u$, $i$-$g$, $i$-$r$, $r$-$u$, and $r$-$g$ band-pairs (where $y$-axis represents shorter wavelengths) are generally negative, and conversely those of $i$-$z$, $r$-$i$, and $r$-$z$ band-pairs (where $y$-axis represents longer wavelengths) are positive.
This means that the total observed flux spectra of these quasars become bluer when they get brighter (``bluer when brighter'' trend), which is a common feature observed in the light curves of the SDSS quasars (see \cite{kok14} for details).

Figure~\ref{fig:comparison_each} shows the relative variable component spectra of the four quasars derived from the linear regression slopes of the flux-flux plots (Table~\ref{regression_table}), along with the polarized component spectra taken from \citet{kis04}.
Figure~\ref{fig:comparison} shows all of these spectra simultaneously to emphasize the similarity of the spectral shape of the variable component spectra between the target quasars.
In Figure~\ref{fig:comparison_each} and Figure~\ref{fig:comparison}, the uncertainty of the variable component spectra in the reference band ($r$-band for B2 1208+32, Ton~202, and 3C323.1, and $i$-band for 4C09.72) is taken to be zero.
Detailed comparisons of the spectral shape of the variable component spectra and the polarized component spectra and discussion about them are given in the next section.

\section{Discussion}
\label{discussion}

\subsection{Notes on the effect of the BLR emission variability}
\label{note1}

In Figure~\ref{fig:comparison_each} and Figure~\ref{fig:comparison}, it is clearly seen that the variable component spectra of the four quasars have very blue power-law like spectral shape throughout the rest-frame near-ultraviolet to optical spectra region, which is as blue as the long-wavelength limit value of the thermal accretion disk prediction $\alpha_{\nu}=1/3$, in good agreement with the previous works on the AGN and quasar variability (e.g., \cite{col99,kok14,rua14,fau15,mac16}; see Section~\ref{intro_variability}).
This means that the variable component spectra are significantly bluer than the polarized component spectra (and thus than the thermal accretion disk model predictions) in the near-ultraviolet spectral region of the four quasars.
Figure~\ref{fig:comparison_each} shows that the discrepancy of the spectral shape between the variable and polarized components (if these spectra are scaled to the same flux level at optical wavelengths) results in about twofold discrepancy of the emitted energy at ultraviolet wavelengths.
As discussed in Section~\ref{intro_goal}, this huge discrepancy strongly suggests that there is fundamental problems in our current understanding of the quasar ultraviolet-optical variability and/or polarization; namely, either (or both) of the two interpretations, one states that the polarized component spectrum should represent the intrinsic accretion disk continuum spectrum (e.g., \cite{kis04,kis08}) and the other states that the variable component spectrum well represents it (e.g., \cite{per06,sch12,kok14,rua14}), is (are) wrong.

It should first be noted that, as mentioned in Section~\ref{intro_variability}, broad-band photometric light curves containing the rest-frame spectral regions of the broad emission lines are unavoidably (more or less) affected by the contamination from the broad line flux variability, although the continuum variability is often larger than the broad line variability at least by a factor of a few (intrinsic Baldwin effect; e.g., \cite{kin90,wil05,bia12,kok14}).
The variable component spectra in Figure~\ref{fig:comparison_each} actually show small deviations from single power-law spectra, which can possibly be attributed to the flux variability of BLR emission; $z$-band of Ton~202 and B2~1208+32 contains the H$\alpha$ line, $i$-band of 4C09.72 and $r$-band of Ton~202 and 3C323.1 contain the H$\beta$ line, and $g$-band contains blends of higher order Balmer lines and the Balmer continuum emission from the BLR, all of which are known to show flux variability as a result of reverberation of the ionizing continuum emission (e.g., \cite{mao93,kor01,kor04,kok14,ede15,fau15}, and references therein).
The dominant BLR emission in $u$-band is Fe\emissiontype{II} pseudo-continuum and Mg\emissiontype{II} emission, but several spectral variability studies show that the variability of these low-ionization emission lines are very weak and they do not significantly contribute to the variable spectral component (see \cite{goa99a,kok14,mod14,sun15}, and references therein).
In any case, it is very difficult to attribute the twofold flux excess in the near-ultraviolet region ($u$ and $g$-band in the observed-frame) in the variable component spectra compared to the polarized component spectra (seen in Figure~\ref{fig:comparison_each}) to the variable BLR emission contamination considering the huge energy required to fill the gap.

More direct observational evidence against the non-accretion disk variable component contamination in near-ultraviolet region is the observed strong inter-band correlation.
The flux-flux plots in Figure~\ref{fig:fluxflux}, especially for 4C09.72, show very strong linear correlation.
Such a strong inter-band correlation is generally observed in AGNs and quasars \citep{sak10,kok15}, and validates the use of the flux gradient method described in Section~\ref{analysis}.
Since the broad line region locates at a few tens or hundreds of light-days (note that our observation period is $\sim$ 200 days in the quasar rest-frame) away from the inner region of the accretion disk in luminous quasars like those analysed in this work (according to the well-established BLR radius-optical luminosity relation; e.g., \cite{ben13}), the observed strong inter-band correlation suggests that the huge flux contribution from the BLR emission is unlikely and that there exists essentially a single dominant variable continuum component throughout the near-ultraviolet to optical wavelengths.

One may think that this power-law like variable component spectra seen in Figure~\ref{fig:comparison_each} and Figure~\ref{fig:comparison} are from the variable optical synchrotron emission.
However, if it was true, the optical polarized component spectra were to be dominated by the intrinsically-polarized synchrotron emission and thus had the same spectral shape with those of the variable component spectra.
The similarity of the variability amplitude and the spectral shape of the variable component observed in the four quasars with the general property of the SDSS quasars, which are mostly radio-quiet quasars (e.g., \cite{ive02}), also suggests that the variable component in these luminous type~1 quasars are related with the accretion disk emission rather than the optical synchrotron emission.
Inversely, it seems to be difficult to explain the observed complex spectral shape of the polarized flux spectra by the optical synchrotron emission contribution (see \cite{kis03,kis04} for more detailed discussion).

\subsection{Possible interpretations of the relationship between variable and polarized spectral components}
\label{note2}

Because there seems to be no explanation for the large flux deficit in the near-ultraviolet region seen in the polarized component spectra of the quasars (compared to the variable component spectra), we may have to consider that the ultraviolet-optical polarization in the luminous type 1 quasars cannot solely explained by the electron (Thomson) scattering, and there may be some unknown (de)polarization mechanisms producing the observed decrease of the polarization degree in the near-ultraviolet wavelengths, considering that the variable flux component spectra should represent the spectral shape of intrinsic accretion disk emission.

On the other hand, it can also be possible that the blue spectral shape of the variable component is not representing the shape of the total flux spectrum coming from the whole surface of the quasar accretion disk.
Generally speaking, a causal argument suggests that the short-term (several days to several tens of days) flux variability observed in AGNs and quasars is caused at the disk radii smaller than several tens of light-days from the central black hole (e.g., \cite{sta04,sun14}).
On the basis of this causal argument and the observational fact of the strong linear inter-band flux-flux correlation of the quasar variability in all time-scale (see e.g., \cite{sak10}), it seems to be natural to consider that the quasar ultra-violet to optical flux variability is always caused by temperature fluctuations within a certain disk radius, which must result in the bluer variable component spectrum than the total flux spectrum from the disk.
Construction of quantitative models based on the above described idea and comparison with observations are beyond the scope of this work, and these will be discussed in the subsequent paper.

Finally, it should also be noted that, although quantitative comparisons with model spectra are currently impossible, there is still a room to consider that the observed polarization properties in the four quasars with $\lambda<$4000\AA~polarization dip may be explained by the intrinsic polarization imprinted in the accretion disk atmosphere. 
By referring to the model calculations by \citet{lao90} (see also \cite{shi98,hsu98,kor99}),  \citet{kis03} and \citet{ant04} stated that, in a certain accretion disk model parameter space, even though the total flux (and thus the variable component) of the accretion disk spectrum has no feature around Balmer edge spectral region, the polarization degree (and thus the polarized component) spectrum can show the decreasing feature at the blueward of Balmer edge due to the increase of the absorption opacity.\footnote{Although these models are generally claimed to suffer from the wrong polarization direction (i.e., polarization direction perpendicular to the disk's rotation axis, contrary to the observations), it is also suggested that absorption opacity effects can in some cases change the polarization position angle to the direction parallel to the disk's rotation axis (Nagirner effect; e.g., \cite{gne78, mat93, ago98}).}
If this is true, our result of the discrepancy of the spectral shape between the variable component and polarized component in the four quasars can be explained so that the variable component spectrum is a scaled copy of the featureless accretion disk continuum, and the polarized component spectrum only represents the intrinsically polarized accretion disk continuum component.

Although above we have listed several implications of the discrepancy between the variable component and polarized component spectra in quasars, it is definitely impossible to conclusively decide either the polarized component or variable component spectrum well represents the intrinsic accretion disk spectrum, or both of the interpretations are invalid, with the currently available data.
The only way to probe the true nature of the relationship between the variable component and the polarized component in the quasar spectra is the examination of the polarimetric variability, which has rarely been investigated for AGNs and quasars (see e.g., \cite{mer06,gas12,afa15a}).
For example, \citet{kis03} noted that Ton~202 showed evidence of slight changes of the polarization degree at $\lambda<4000$\AA\ dip region within a year time-scale.
This may indicate the presence of the fast-moving absorption material responsible for the Balmer-edge feature in the edge-on trajectory between the inner accretion disk region and the scatterer, which must be confirmed by further spectropolarimetric follow-up observations.
Future intensive photometric and/or spectroscopic polarimetric monitoring observations will clarify the causes of the discrepancy between the spectral shapes of the polarized component and variable component in luminous type~1 non-blazar quasars.

\section{Summary and conclusions}
\label{summary}

In the literature, it is suggested that the ultraviolet-optical polarized flux spectra of the luminous type~1 quasars are representing the intrinsic spectral shape of the quasar accretion disk emission, and several quasars are showing Balmer edge features (specifically, decrease of the polarized flux at $\lambda<4000$\AA) in their polarized flux spectra which can be interpreted as the imprint of the opacity effect in the accretion disk atmosphere (e.g.,  \cite{kis03,ant04,kis04,kis08,hu12}).
On the other hand, it is also assumed in several previous works that the ultraviolet-optical variable component spectra of quasars are the good indicator of the intrinsic spectral shape of the quasar accretion disk emission (e.g., \cite{per06,sch12,kok14,rua14}).
In this work, we examined the consistency of the above mentioned assumptions through the investigation of whether the variable component spectra have the same spectral shape with the polarized flux component spectra in a sample of four $\lambda<4000$\AA~polarization-decreasing quasars spectropolarimetrically confirmed by \citet{kis04} (4C09.72, 3C323.1, Ton~202, and B2~1208+32).
The result is negative, in that the variable component spectra are significantly bluer compared to the polarized flux component spectra especially in the near-ultraviolet spectral region; the variable component spectra of these quasars are confirmed to be well represented by a single power-low component with $\alpha_{\nu} \sim 1/3$ through the rest-frame ultraviolet-optical wavelength range, resulting in the twofold excess of the emitted energy at ultraviolet wavelengths compared to the polarized flux spectra.
Although it is impossible to decide which (both) of the two assumptions is (are) invalid only from the currently available observational constraints, we can at least say that this discrepancy in the spectral shape implies either (1) the decrease of polarization degree in the rest-frame ultraviolet wavelengths is not indicating the Balmer absorption edge feature but is induced by some unknown (de)polarization mechanisms, or (2) the ultraviolet-optical flux variability is occurring preferentially at the hot inner radii of the accretion disk and thus the variable component spectra do not reflect the whole accretion disk emission.
Future photometric and/or spectroscopic polarimetric monitoring observations will be useful to clarify the causes of this discrepancy between the spectral shapes of the polarized component and variable component, and consequently the true nature of the accretion disk emission in the luminous type~1 non-blazar quasars.

\begin{ack}

We thank Yuki Sarugaku for his support during the KWFC queue mode observations.
We are grateful to all the staff in the Kiso Observatory for their efforts to maintain the observation system.
We thank the referee, Makoto Kishimoto, for careful reading of our manuscript and for giving useful comments. 
This work was supported by JSPS KAKENHI Grant Number 15J10324.

This research has made use of NASA's Astrophysics Data System Bibliographic Services.
This research has made use of the NASA/IPAC Extragalactic Database (NED), which is operated by the Jet Propulsion Laboratory, California Institute of Technology, under contract with the National Aeronautics and Space Administration.
Funding for SDSS-III has been provided by the Alfred P. Sloan Foundation, the Participating Institutions, the National Science Foundation, and the U.S. Department of Energy Office of Science. The SDSS-III web site is http://www.sdss3.org/.
SDSS-III is managed by the Astrophysical Research Consortium for the Participating Institutions of the SDSS-III Collaboration including the University of Arizona, the Brazilian Participation Group, Brookhaven National Laboratory, Carnegie Mellon University, University of Florida, the French Participation Group, the German Participation Group, Harvard University, the Instituto de Astrofisica de Canarias, the Michigan State/Notre Dame/JINA Participation Group, Johns Hopkins University, Lawrence Berkeley National Laboratory, Max Planck Institute for Astrophysics, Max Planck Institute for Extraterrestrial Physics, New Mexico State University, New York University, Ohio State University, Pennsylvania State University, University of Portsmouth, Princeton University, the Spanish Participation Group, University of Tokyo, University of Utah, Vanderbilt University, University of Virginia, University of Washington, and Yale University.
\end{ack}

%%Bib
\bibliography{quasar_polarimetric_variability_kiso}

\end{document}